\documentclass{article}
\def\beq{\begin{equation}}
\def\eeq{\end{equation}}
\def\bea{\begin{eqnarray}}
\def\eea{\end{eqnarray}}
\title{}
\begin{document}
\title{Poincar\'e Invariance, Cluster Properties, and Particle Production}
\author{W. N. Polyzou\thanks{\textit{E-mail address:} 
polyzou@uiowa.edu} \thanks{
This work Supported by the U.S. Department of
Energy, Nuclear Physics Division, contract
DE-FG02-86ER40286.} \thanks{Contributed talk to the XVIII 
European Conference on Few-Body Problems}
\\ \it The University of Iowa}

\maketitle
\begin{abstract}
A method is presented for constructing a class of Poincar\'e invariant
quantum mechanical models of systems of a finite number of degrees of
freedom that satisfy cluster separability, the spectral condition,
but do not conserve particle number.  The class of models includes the
relativistic Lee model \cite{fcwp} and relativistic isobar models.
\end{abstract}

\bigskip

Any reasonable quantum theory of strongly interacting particles should
be Poincar\'e invariant, and should satisfy cluster properties and the
spectral condition.  These conditions are physical requirements.  In
addition, the theory should be able to model reactions that change
particle number.  These conditions require the existence of a unitary
representation of the Poincar\'e group $\hat{U}(\Lambda,a)$
\cite{wigner} on the model Hilbert space ${\cal H}$ that satisfies a
cluster separability condition and the spectral condition.

While these requirements are dictated by physics, it is difficult to
find well-defined mathematical models for realistic systems that are
consistent with all of these properties.  Relativistic quantum theory
of $N$-particles\cite{fcwp}\cite{wp2} is a mathematically well-defined
theory where these conditions can be satisfied.  A mathematically
well-defined theory is required to formulate high-precision ab-initio
calculations, which are important for strongly interacting systems.

This talk explains how to extend the relativistic quantum theory of
$N$-particles \cite{fcwp}\cite{wp2} to treat a class of models with 
particle production.  This class of models has a
bounded number of degrees of freedom and includes the relativistic Lee
model \cite{fcwp} and relativistic isobar models.

The key observation is that mass is not superselected in relativistic
quantum theory.  To contrast this with non-relativistic quantum
theory, consider a reaction involving a fixed number of constituent
particles, where a bound target is broken up into a number of
subsystems.  In this reaction, the mass and number of particles in the
initial and final states is different.  This reaction cannot be
consistently modeled with a non-relativistic quantum theory because it
violates Galilean invariance if the Galilean and inertial masses are
identified.

The difference between the above reaction and a general production
reaction is the presence of a conserved number of constituent
particles, which limits the number of degrees of freedom.  Motivated
by this observation, this talk considers production reactions that
have a set of conserved quantum numbers which are associated with
particles called ``atomic partons''.  Additional bare particles with
{\it composite} atomic-parton quantum numbers are introduced.  The
model Hilbert space is a direct sum of tensor products of bare and
atomic particle subspaces with the same atomic-parton numbers.  The
dynamics is restricted to conserve atomic-parton number.

For example, a model where the atomic partons are a nucleon and two
pions could have a bare $\Delta$, with the quantum numbers of a
nucleon and a pion, and a bare $\rho$, with the quantum numbers of two
pions.  Introducing bare particles does not mean that these particles
exist as stable particles. Stable particles correspond to point
eigenstates of the mass operator.  This talk investigates the
modifications of the fixed $N$-construction \cite{fcwp}\cite{wp2}
which are needed to treat this type of particle production.
 
In the above example the conserved atomic-parton numbers are ${\cal N}:=
\{ N_N, N_\pi\}$.  The Hilbert space, ${\cal H}_{\cal N}$, is a direct
sum of tensor products of irreducible representation spaces of the
Poincar\'e group, where the particles in each tensor product have
atomic parton number ${\cal N}$
\[
{\cal H}_{\cal N}  = \oplus_{i} (\otimes {\cal H}_{a_i});\, \mbox{i.e.}
\quad 
{\cal H}_{\{1,1\}}  = ({\cal H}_{N}\otimes {\cal H}_{\pi})\oplus{\cal H}_{\Delta}. 
\]
There is a natural partial ordering on the quantum numbers ${\cal N}$
given by ${\cal N}\leq {\cal N}'$ if $N_i\leq N_i'$ for each type,
$i$, of atomic parton.  The space ${\cal H}_{\cal N}$ has a natural
non-interacting representation of the Poincar\'e group,
$\hat{U}_{0}(\Lambda ,a)$
 
For each partition, $b$, of the atomic partons into disjoint
equivalence classes, $b_i$, there is a factorization of the model
Hilbert space, ${\cal H}_{\cal N}$, into a direct sum of a tensor
product, ${\cal H}_b$, of $n_b$ proper subsystem Hilbert spaces, ${\cal
H}_{b_i}$, with ${\cal N}_i < {\cal N}$, and a residual space, ${\cal
H}^b$, corresponding to bare particles with quantum numbers in more
than one equivalence class:
\[
{\cal H}_{{\cal N}} = {\cal H}_b \oplus {\cal H}^b
\qquad
{\cal H}_b := \otimes_{b_i \in b}  {\cal H}_{b_i} . 
\] 

The difference between the construction outlined in this talk and the
construction for fixed $N$ in \cite{fcwp}\cite{wp2} involves the
modifications that are necessary to treat the residual space ${\cal
H}^b$.  The asymptotic dynamics and translation operators associated
with partition $b$ are needed to formulate cluster properties.  On
${\cal H}={\cal H}_b \oplus {\cal H}^b$ these operators can be
expressed in terms of subsystem representations $\hat{U}_{b_i}(\Lambda
,a)$ as:
\beq
\hat{U}_b (\Lambda ,a) := 
\left ( 
\begin{array}{cc}
\otimes \hat{U}_{b_i}(I ,y_i) &0 \\
0& \hat{I}^b
\end{array}
\right )
\qquad
\hat{T}_b (y_1, \cdots , y_k ) := 
\left ( 
\begin{array}{cc}
\otimes_i \hat{U}_{b_i}(I ,y_i) &0 \\
0& \hat{I}^b
\end{array}
\right ).
\label{eq:AA}
\eeq
The projection on ${\cal H}_b$, 
\[
\hat{\Pi}_b  := 
\left ( 
\begin{array}{cc}
\hat{I} &0 \\
0& 0
\end{array}
\right ) ,
\]
is used to define a modified formulation of the cluster property
\beq
\lim_{\mbox{min}(y_i-y_j)^2 \to \infty}  
\Vert [\hat{U}(\Lambda ,x) - \hat{U}_{b}(\Lambda ,x)]
\hat{T}_{b}(y_1, \cdots, y_k ) \hat{\Pi}_b \vert \psi 
\rangle \Vert =0 .
\label{eq:clus}
\eeq
This condition requires more than a range condition on the interaction;
it also requires that part of the operator that survives asymptotically 
is a tensor product of proper-subsystem representations \cite{wp2}.  

In this formulation ${\cal N}$-body interactions, $\hat{V}_{\cal N}$,
satisfy 
\[
\lim_{\mbox{min}(y_i-y_j)^2 \to \infty} 
\Vert \hat{V}_{{\cal N}} 
\hat{T}_{b}(y_1 ,\cdots y_k ) \hat{\Pi}_b \vert \psi \rangle 
\Vert =0 \quad \forall \, b .
\]
The dynamics of all proper subsystems determine the
dynamics of the full system up to an ${\cal N}$-body interaction.

The construction of a two atomic parton dynamics uses {\it Wigner's
forms of the dynamics} \cite{wp1}\cite{wp2}; Clebsch-Gordan
coefficients of the Poincar\'e group are used to construct an
irreducible basis for the non-interacting system.  Interactions are
added to the non-interacting mass that are block-diagonal in the
non-interacting spin, $\hat{j}_0$, and commute with and are
independent of the commuting functions, $\hat{F}_0$, of the Poincar\'e
generators that label vectors in irreducible subspaces.  It follows
that simultaneous eigenstates of the mass, $\hat{M}$, $\hat{j}_0$, and
$\hat{F}_0$ define an irreducible basis for an interacting
representation of the Poincar\'e group.  In this representation the
mass eigenvalue and non-interacting spin are the new Casimir
operators.

The many-body construction is recursive.  The representations
$\hat{U}_b (\Lambda ,a)$ are constructed from tensor products of
proper subsystem, $({\cal N}_i < {\cal N})$, representations.  If
$\hat{U}_b (\Lambda ,a)$ is decomposed into irreducible
representations, the spin ,$\hat{j}$, and the vector operators, $\hat{F}$,
depend on the interaction.  If the order of constructing irreducible
representations and adding interactions is reversed, the result is a
mass operator, $\bar{M}_b$, that commutes with $\hat{j}_0$, and
$\hat{F}_0$, and is independent of $\hat{F}_0$.  Simultaneous
eigenstates of $\bar{M}_b$, $\hat{j}_0$, and $\hat{F}_0$ transform
irreducibly and define a representation, $\bar{U}_b (\Lambda ,a)$, of
the Poincar\'e group.  The representations $\bar{U}_b (\Lambda ,a)$
and $\hat{U}_b (\Lambda ,a)$ are related by a unitary scattering
equivalence $\hat{B}_b$ \cite{fcwp}\cite{wp1} of the form:

\beq
\hat{B}_b  := 
\left ( 
\begin{array}{cc}
\hat{A}_b &0 \\
0& \hat{I}^b
\end{array}
\right ) 
\qquad
\bar{U}_b (\Lambda ,a)= \hat{B}_b \hat{U}_b (\Lambda ,a) \hat{B}_b^{\dagger}
\label{eq:bar}
\eeq
which satisfies
\[
\lim_{t \to \pm\infty}  
\Vert [(\hat{B}_b -\hat{I}) \hat{\Pi}_0  \hat{U}_0(I ,t) 
\vert \psi \rangle \Vert =0,
\]
where $\hat{\Pi}_0$ is the projector on the ${\cal N}$-atomic parton
subspace of ${\cal H}$.  For systems of more than three atomic
partons, the freedom to choose the operators $\hat{B}_b$ must be
utilized to ensure that the $\hat{B}_b$ are also consistent with
cluster properties.  The techniques used the $N$-particle
case\cite{fcwp}\cite{wp1} can be applied to the modified operators
(\ref{eq:AA}) and (\ref{eq:bar}).

All interactions appear in the $\bar{U}_b(\Lambda,a)$ as $b$ ranges
over all partitions with at least two equivalence classes, except the
${\cal N}$-body interactions.  The mass operators $\bar{M}_b$ for the
different $\bar{U}_b (\Lambda ,a)$ can be combined, using an Ursell
cluster expansion, to define
\[
\bar{M} := \sum_b (-)^{n_b} (n_b-1)!
\hat{B}_b \hat{M}_b \hat{B}_b^{\dagger}, 
\]
which commutes with the non-interacting ${\cal
N}$-body spin $\hat{j}_0$ and the non-interacting $\hat{F}_0$.
Simultaneous eigenstates of $\bar{M}$, $\hat{j}_0$, $\hat{F}_0$ define a
basis that transforms irreducibly and thus a representation
$\bar{U}(\Lambda ,a)$ of the ${\cal N}$-body dynamics.

The representation $\bar{U}(\Lambda ,a)$ fails to satisfy the cluster property
(\ref{eq:clus}) because of the transformations $\hat{B}_b$ appearing in 
Eq. (\ref{eq:bar}).   This can be remedied by defining 
\[
\hat{B}= {1 - i \hat{\beta} \over 1+i \hat{\beta}} 
\qquad
\hat{\beta} = \sum_b (-1)^{n_b} (n_b-1)!  \hat{\beta}_b
\]
\[
\hat{\beta}_b := i 
{ \hat{B}_b -\hat{I}  \over \hat{B}_b+\hat{I}} .
\]
When the Cayley transforms,  $\hat{\beta}_b$, are bounded the 
operator $\hat{B}$ is a scattering equivalence and the representation   
\[
\hat{U}(\Lambda ,a) := \hat{B}^{\dagger} \bar{U}(\Lambda ,a) \hat{B}
\]
satisfies cluster properties (\ref{eq:clus}) for suitable short-ranged 
\cite{wp2} interactions.

This construction shows how the method for constructing
relativistic quantum theory of a fixed number of particles can be
extended to treat a class reactions with particle production.
Conservation of atomic parton number allowed the separation of the
difficulties associated with particle production from the difficulties
associated with having an infinite number of degrees of freedom.


\begin{thebibliography}{99}

\bibitem{fcwp} F. Coester and W.N. Polyzou: Phys. Rev. D {\bf 26} 1348 (1982)

\bibitem{wigner} E. P. Wigner: Ann. Math. {\bf 40}, 141 (1939)

\bibitem{wp2} W.N. Polyzou: J. Math. Phys. In press; nucl-th/0201013

\bibitem{wp1} W. Polyzou: Ann. Phys.  193, 367(1989)

\end{thebibliography}
\end{document}